\documentclass[aps, prl, twocolumn, letterpaper, nofootinbib, superscriptaddress, balancelastpage, longbibliography]{revtex4-1}

\usepackage{graphicx}
\usepackage{dcolumn}
\usepackage{amsmath, amsthm, mathrsfs, amsfonts, dsfont}
\usepackage{url}
\usepackage{xcolor}
\usepackage[colorlinks,citecolor=blue,urlcolor=blue,linkcolor=blue]{hyperref}
\usepackage{bm, bbold, xcolor, slashed}
\usepackage{easy-todo}
\usepackage{stmaryrd}
\usepackage{braket}

\def\caravel{{\textsc{Caravel}}} 
\def\xact{{\textsc{xAct}}} 

\begin{document}

\preprint{APS/123-QED}

\title{First Look at Quartic-in-Spin Binary Dynamics at Third Post-Minkowskian Order}

\author{Dogan Akpinar}
\email{dogan.akpinar@ed.ac.uk} 
\affiliation{Higgs Centre for Theoretical Physics, School of Physics and
Astronomy, University of Edinburgh, EH9 3FD, UK} 
\author{Fernando Febres Cordero}
\email{ffebres@hep.fsu.edu} 
\affiliation{Physics Department, Florida State University, Tallahassee, FL
32306-4350,USA} 
\author{Manfred Kraus}
\email{mkraus@fisica.unam.mx} 
\affiliation{Departamento de F\'{i}sica Te\'{o}rica, Instituto de F\'{i}sica,
\\ Universidad Nacional Aut\'{o}noma de M\'{e}xico, Cd. de M\'{e}xico C.P.
04510, M\'{e}xico} 
\author{Alexander Smirnov}
\email{asmirnov@srcc.msu.ru}
\affiliation{Research Computing Center, Moscow State University, Moscow, Russia}
\affiliation{Moscow Center for Fundamental and Applied Mathematics, Moscow, Russia}
\author{Mao Zeng}
\email{mao.zeng@ed.ac.uk} 
\affiliation{Higgs Centre for Theoretical Physics, School of Physics and
Astronomy, University of Edinburgh, EH9 3FD, UK}%

\date{\today}

\begin{abstract}
  We compute the conservative and radiation-reaction contributions to
  classical observables in the gravitational scattering between a
  spinning and a spinless black hole to the fourth order in spin and
  third order in the gravitational constant.  The conservative results
  are obtained from two-loop amplitudes for the scattering process of
  a massive scalar with a massive spin-$s$ field $(s=0, 1, 2)$
  minimally coupled to gravity, employing the recently introduced spin
  interpolation method to resolve all spin-Casimir terms. The two-loop
  amplitude exhibits a spin-shift symmetry in both probe limits, which
  we conjecture to be a sign of yet unknown integrability of Kerr
  orbits through the quartic order in spin and to all orders in the
  gravitational constant. We obtain the radial action from the finite
  part of the amplitude and use it to compute classical observables,
  including the impulse and spin kick.
  This is done using the recently introduced covariant Dirac brackets,
  which allow for the computation of classical scattering observables
  for general (nonaligned) spin configurations.  Finally, employing
  the radiation-reaction amplitude proposed by Alessio and Di Vecchia,
  together with the Dirac brackets, we obtain radiation-reaction
  contributions to observables at all orders in spin and beyond the
  aligned-spin limit.
  We find agreement with known  
  results up to the quadratic order in spin for both conservative
  and radiation-reaction contributions. Our results advance the state
  of the art in the understanding of spinning binary dynamics in
  general relativity and demonstrate the power and simplicity of the
  Dirac bracket formalism for relating scattering amplitudes to
  classical observables.
\end{abstract}

\maketitle

\textbf{\textit{Introduction---}}The LIGO-Virgo-KAGRA collaboration
is steadily advancing our understanding of gravitational wave (GW)
physics.  During the current O4 run it is expected that the number of
black hole and neutron star merger events detected will more than
double. Furthermore, the next-generation GW
observatories~\cite{Punturo:2010zz,LISA:2017pwj,Reitze:2019iox} will
greatly increase detector sensitivity and scientific reach
\cite{Borhanian:2022czq}. The data will challenge existing
theoretical frameworks for producing correspondingly precise
predictions \cite{Purrer:2019jcp}. Indeed, progress in the field
hinges from detailed comparisons between experimental measurements and
theoretical predictions, which allows for the inference of parameters of the detected events,
for example the masses and spins of the interacting
compact objects.

Recently it has been shown that the dynamics of binary compact systems
interacting gravitationally have a natural description within the
context of quantum field theory (QFT). Most notably, the so-called
\textit{post-Minkowskian} (PM) expansion mirrors the standard
relativistic amplitude computation in QFT, where one expands
perturbatively in a single small coupling while keeping full
dependence on the velocity. In fact, relativistic QFT techniques have
emerged as a new workhorse for computing PM corrections to the dynamics
of spinless binaries through on-shell scattering amplitude methods,
e.g.,\ in Refs.~\cite{Cheung:2018wkq, Kosower:2018adc, Bern:2019nnu,
  Bern:2019crd, Cristofoli:2019neg, Bjerrum-Bohr:2019kec,
  Brandhuber:2021eyq, Bern:2021dqo, Bern:2021yeh, Damgaard:2023ttc},
as well as worldline methods, e.g.,\ in Refs.~\cite{Kalin:2020mvi,
  Kalin:2020fhe, kalin:2022hph, Dlapa:2023hsl, Dlapa:2021npj,
  Mogull:2020sak, Jakobsen:2021smu, Jakobsen:2022psy,
  Jakobsen:2023oow, Driesse:2024xad, Driesse:2024feo}.

The effects of spin in compact bodies adds another level of
complexity.  Incorporating them into the PM expansion has required
many novel developments, both in scattering amplitudes and worldline
approaches; see e.g.,\ Refs.~\cite{Bini:2017xzy, Bini:2018ywr,
  Vines:2017hyw, Vines:2018gqi, Guevara:2017csg, Guevara:2018wpp,
  Chung:2018kqs, Arkani-Hamed:2019ymq, Guevara:2019fsj, Chung:2019duq,
  Damgaard:2019lfh, Aoude:2020onz, Chung:2020rrz, Guevara:2020xjx,
  Bern:2020buy, Kosmopoulos:2021zoq, Chen:2021qkk,
  FebresCordero:2022jts, Bern:2022kto, Bern:2023ity, Menezes:2022tcs,
  Riva:2022fru, Damgaard:2022jem, Aoude:2022thd, Aoude:2022trd,
  Bautista:2022wjf, Gonzo:2023goe, Aoude:2023vdk, Lindwasser:2023zwo,
  Brandhuber:2023hhl, DeAngelis:2023lvf, Aoude:2023dui,
  Bohnenblust:2023qmy, Gatica:2024mur, Cristofoli:2021jas,
  Luna:2023uwd, Gatica:2023iws, Liu:2021zxr, Jakobsen:2021lvp,
  Jakobsen:2021zvh, Jakobsen:2022fcj, Jakobsen:2022zsx,
  Jakobsen:2023ndj, Jakobsen:2023hig, Heissenberg:2023uvo,
  Lindwasser:2023dcv, Bautista:2023sdf, Cangemi:2023ysz,
  Brandhuber:2024bnz, Chen:2024mmm, Bhattacharyya:2024kxj,
  Alaverdian:2024spu, Brandhuber:2024qdn, Brandhuber:2024lgl,
  Akpinar:2024meg, Bohnenblust:2024hkw, Haddad:2024ebn,
  Bonocore:2024uxk}. In this way, spectacular results have been
achieved at $O(G)$ and $O(G^2)$, corresponding to tree-level and
one-loop amplitude calculations, respectively.  Furthermore, two-loop
results, i.e., $\mathcal{O}(G^3)$, from scattering amplitudes for
observables at the second order in spin, for systems with one spinless
and one spinning black hole, have been calculated in
~\cite{FebresCordero:2022jts, Akpinar:2024meg}. Worldline techniques
have also been developed for the computation of two-loop corrections
to observables with two powers of spin~\cite{Jakobsen:2022fcj},
three-loop corrections [$\mathcal{O}(G^4)$] to linear-in-spin
observables~\cite{Jakobsen:2023ndj}, and, more recently, the framework
has been extended to obtain results to fourth order in spin at the
one-loop level~\cite{Ben-Shahar:2023djm,Haddad:2024ebn}.

Going beyond the second order in spin at two or more loops has been a
major challenge. The massive spinor helicity
formalism~\cite{Arkani-Hamed:2017jhn} with all-order spin
exponentiation suffers from the appearance of arbitrarily high-rank
tensor integrals. The arbitrary-spin Lagrangian
formalism~\cite{Bern:2020buy} has the strength that it encompasses
subtle effects such as spin magnitude changes
\cite{Alaverdian:2024spu}, but requires careful choices of Wilson
coefficients \cite{Bern:2023ity}.  To make progress, we adopt the
approach of using scattering amplitudes involving fixed spin-$s$
fields, minimally coupled to gravity, to determine spin effects in
classical two-body interactions up to 2$s$ orders in the spin
multipole expansion~\cite{Ross:2007zza, Holstein:2008sw,
  Vaidya:2014kza, Maybee:2019jus} specifically for the case of Kerr
black holes. However such an approach requires subtle considerations.
When performing calculations this way, an ambiguity appears in
spin-Casimir terms, like $S^2$, starting at one loop.  This can be
seen by the following power-counting argument. 
In the classical limit, the momentum $q$ of the exchanged gravitons
scales as $\mathcal{O}(\hbar)$ while the spin can become large. Then
we see that, for fixed small values of $s$ and employing natural units
($\hbar=1$), the spin-Casimir $q^2 S^2 = -q^2 s(s+1)$ inevitably mixes
with $q^2\sim\mathcal{O}(\hbar^2)$ quantum corrections to the
spin-independent part of the amplitude when performing an expansion in
$q$.

In this Letter, we compute for the first time the classical amplitude
for the scattering of a spinning black hole with a spinless one to
fourth order in spin and to third order in the gravitational coupling,
i.e.\ to $\mathcal{O}(G^3S^4)$. Considering that for the Kerr black 
hole $|S|<Gm^2$, this physically can be counted as up to the seventh 
post-Minkowskian order (see Ref.~\cite{Rettegno:2023ghr}).  We make use 
of the \textit{spin interpolation} method~\cite{Akpinar:2024meg}, 
recently introduced by
some of the present authors and M.~S. Ruf, to systematically resolve
ambiguities between quantum corrections of the scattering amplitudes
and classical spin-Casimir terms, without resorting to arbitrary-spin
formalisms that are challenging beyond one loop.
In our amplitude we observe a spin-shift symmetry, previously
established at one loop ~\cite{Aoude:2022trd,Bern:2022kto}, as well as
two loops truncated at lower orders in spin \cite{Akpinar:2024meg}, up
to the fourth order in spin in the limit when either the spinless or
spinning black hole becomes a probe object. We also establish the
equivalence of the radial action, obtained from the two-loop
amplitudes via the amplitude-action relation~\cite{Bern:2021dqo,
  Bern:2021yeh, Bern:2024adl}, to known probe-limit results from
direct classical calculations \cite{Gonzo:2024zxo}. Then, we use the
covariant Dirac brackets for spinning two-body dynamics recently
introduced by Gonzo and Shi~\cite{Gonzo:2024zxo} (see also
Refs.~\cite{Kim:2024grz, Kim:2024svw, Kim:2025hpn} for a related
Poisson bracket approach), to compute the impulse and spin kick
observables for general (nonaligned) spin configurations. This
simplifies the calculation compared with a direct application of the
Kosower-Maybee-O'Connell formalism \cite{Kosower:2018adc} generalized
to include spin~\cite{Maybee:2019jus}. Finally, we combine our results
with the known radiation-reaction results of
Ref.~\cite{Alessio:2022kwv} and find cancellation of a high-energy
logarithmic divergence through quartic order in spin. We compute, for
the first time, the radiation-reaction contribution to the observables
mentioned before to all orders in spin beyond the aligned-spin limit
using the covariant Dirac brackets.

\textbf{\textit{Scattering amplitudes---}}As we are interested in computing observables through fourth order
in spin, we consider the following Lagrangian
\begin{multline}
 \frac{\mathcal{L}}{\sqrt{-g}} = -\frac{2R}{\kappa^2} + \frac{1}{2}g^{\mu\nu}(\partial_\mu\phi)(\partial_\nu\phi) - \frac{1}{2}m_2^2\phi^2 + \mathcal{L}_{m_1}\;,
\end{multline}
where we work in the mostly minus metric signature,
$\kappa = \sqrt{32\pi G}$, $g_{\mu\nu}$ and $g$ is the metric and its
determinant, and $R$ is the Ricci scalar.  $\mathcal{L}_{m_1}$ denotes
the Lagrangian of an additional minimally coupled massive field (with
mass $m_1$), which can be either a scalar field $\Phi$
$(s=0)$~\cite{Klein:1926tv,Fock:1926fj,Gordon:1926emj}, a Proca field
$V^\mu$ $(s=1)$~\cite{Proca:1936fbw}, or a Fierz-Pauli field
$H^{\mu\nu}$ $(s=2)$~\cite{Fierz:1939ix} (see
Ref.~\cite{Chiodaroli:2021eug} for more discussions). For the latter
we have
\allowdisplaybreaks{
\begin{align}
 \mathcal{L}_H &= g^{\mu\nu}g^{\alpha\beta}g^{\omega\sigma}\Big\{\frac{1}{2}(\nabla_\omega H_{\mu\alpha})(\nabla_\sigma H_{\nu\beta}) \nonumber\\
 & - (\nabla_\mu H_{\alpha\omega})(\nabla_\beta H_{\nu\sigma}) + (\nabla_\mu
H_{\nu\alpha})(\nabla_\beta H_{\omega\sigma}) \label{eq:FPlag}\\
 &- \frac{1}{2}(\nabla_\mu H_{\alpha\beta})(\nabla_\nu H_{\omega\sigma}) \Big\} \nonumber \\
 &- \frac{1}{2}m_1^2 g^{\mu\nu}g^{\alpha\beta}\big( H_{\mu\alpha}H_{\nu\beta} - H_{\mu\nu}H_{\alpha\beta}\big)\;,\nonumber
\end{align}}%
where $\nabla_\mu$ is the covariant derivative defined with a Christoffel
connection. We expand around a flat background metric
$g_{\mu\nu} = \eta_{\mu\nu} + \kappa h_{\mu\nu}$, where $h_{\mu\nu}$ is the
graviton field.  
All scattering amplitudes are computed with the multiloop numerical unitarity
method~\cite{Ita:2015tya,Abreu:2017idw,Abreu:2017xsl,Abreu:2017hqn,Abreu:2018jgq}
as implemented in the \caravel{} framework~\cite{Abreu:2020xvt}.
In particular, \caravel{} has been employed in multiple applications to quantum and classical
gravity calculations~\cite{Abreu:2020lyk,FebresCordero:2022jts,Bohnenblust:2023qmy,Akpinar:2024meg},
and in this work we have extended it to include massive spin-$2$ bosons according to
the Lagrangian in equation~\eqref{eq:FPlag}.
Corresponding Feynman rules were derived with the \xact{}
package~\cite{Brizuela:2008ra,Nutma:2013zea} and external polarization states
have been taken from~\cite{Bonifacio:2017nnt}. 
As an important validation of our implementation, we have found that
all three-, four- and five-point tree-level amplitudes involved in the
unitarity cuts of the two-loop amplitudes agree numerically with those
obtained via the double copy~\cite{Kawai:1985xq, Bern:2008qj,
  Bern:2010ue, Cachazo:2013gna, Cachazo:2013hca, Edison:2020ehu} with
dimensional reduction to massive amplitudes following
e.g.,~\cite{Bern:2019crd, Johansson:2019dnu, Bautista:2019evw,
  Chiodaroli:2021eug}.  Furthermore, we have found agreement with
analytic constructions of the relevant $2 \to 2$ tree-level and
one-loop amplitudes.
Additional technical details about our calculations can be found
in~\cite{Akpinar:2024meg,Abreu:2020xvt}.
The scattering processes that we consider are
\begin{equation}
	X(-p_1, \varepsilon_1) + \phi(-p_2) \rightarrow \phi(p_3) + X(p_4,\varepsilon^\star_4)\;,
\label{eq:process}
\end{equation}
where $X = \{\Phi, V^\mu, H^{\mu \nu}\}$ and $\varepsilon_1,\varepsilon^\star_4$
are the polarizations associated to the massive spinning particle. We have the
on-shell conditions $p_1^2 = p_4^2 = m_1^2, \ p_2^2 = p_3^2 = m_2^2$ and the
transverse conditions $p_1 \cdot \varepsilon_1 = p_4 \cdot \varepsilon^\star_4 =
0$. The momenta are parametrized~\cite{Landshoff:1969yyn,Parra-Martinez:2020dzs}
as 
\begin{equation}
\begin{split}
	p_1 & = -\bar{p}_1 + q/2\;, \quad p_4 = \bar{p}_1 + q/2\;,\\
	p_2 & = -\bar{p}_2 - q/2\;, \quad p_3 = \bar{p}_2 - q/2\;,
\end{split}
\end{equation}
with $\bar{p}_i = \bar{m}_i u_i$, $\bar{p}_i \cdot q = 0$, the $u_i$ are unit
vectors and $\bar{m}_i^2=m_i^2-q^2/4$. We further define 
$y = u_1 \cdot u_2$, which is a measure of the rapidity difference between the
two massive particles that scatter.

It is convenient to introduce a form factor decomposition for the scattering
amplitudes that involve massive spinning fields.  For the amplitudes involving
$X = V^\mu$ we use the decomposition in Ref.~\cite{Akpinar:2024meg}, while for
those involving $X = H^{\mu \nu}$ we write 
\begin{equation}
\mathcal{M} = \sum_{n =1}^{9}M_n~\varepsilon^\star_{4\mu\nu}
 T_n^{\mu\nu\alpha\beta}\varepsilon_{1\alpha\beta}\;,
\label{eq:FFdec}
\end{equation}
with
\allowdisplaybreaks
\begin{equation}
\begin{split}
&T_1^{\mu \nu \alpha \beta} = \eta^{\mu \alpha}\eta^{\nu\beta}\;, \\
&T_2^{\mu \nu \alpha \beta} = \bar{p}_2^\mu\bar{p}_2^\nu\bar{p}_2^\alpha\bar{p}_2^\beta\;, \\
&T_3^{\mu \nu \alpha \beta} = \bar{p}_2^\mu\eta^{\nu\alpha}\bar{p}_2^\beta\;, \\
&T_4^{\mu \nu \alpha \beta} = 2\bar{p}_2^\nu\bar{p}_2^\alpha \bar{p}_2^{[\mu} q_{\phantom{2}}^{\beta]}\;, \\
&T_5^{\mu \nu \alpha \beta} = 2\eta^{\nu \alpha} \bar{p}_2^{[\mu} q_{\phantom{2}}^{\beta]}\;, \\
\end{split}
\hspace{0.2cm}
\begin{split}
&T_6^{\mu \nu \alpha \beta} = q^\mu \bar{p}_2^\nu\bar{p}_2^\alpha q^\nu\;, \\
&T_7^{\mu \nu \alpha \beta} = q^\mu \eta^{\nu \alpha} q^\beta\;, \\
&T_8^{\mu \nu \alpha \beta} = 2q^\mu \bar{p}_2^{[\nu} q_{\phantom{2}}^{\alpha]}q^\beta\;, \\
&T_9^{\mu \nu \alpha \beta} = q^\mu q^\nu q^\alpha q^\beta\;,
\end{split}
\label{eq:TensorStructures}
\end{equation}
where we use $x^{[\mu} y^{\nu]} = (x^\mu y^\nu - x^\nu y^\mu) / 2$. In
practice, there is one more crossing symmetric tensor which, however,
is evanescent, i.e.\ linearly dependent on the previous structures in
$4$ space-time dimensions. The form factors $M_n$ in \eqref{eq:FFdec}
contain all the theory-dependent information. They are given as linear
combinations of 2781 Feynman integrals with various propagator
structures and including irreducible scalar products in the
numerator. The integrals are expanded in the soft region and reduced
to master integrals via integration-by-parts (IBP)
relations~\cite{Chetyrkin:1981qh, Laporta:2000dsw}. To isolate
conservative dynamics, the master integrals are evaluated in the
potential region following Ref.~\cite{Parra-Martinez:2020dzs}. The IBP
reduction was performed using a private version of the \textsc{FIRE}
program which builds upon
Refs.~\cite{Smirnov:2019qkx,Smirnov:2023yhb,Smirnov:2024onl}; this was
previously used for four-loop spinless binary dynamics calculations in
toy model theories \cite{Bern:2023ccb, Bern:2024adl}.  We used an
approach based on modular arithmetics~\cite{vonManteuffel:2014ixa,
  Peraro:2016wsq} running multiple reductions with different values of
variables.  It was followed by a functional reconstruction procedure
\cite{Peraro:2016wsq} based on Thiele's interpolation formula for the
first variable and on a balanced Zippel approach for the second
variable \cite{Smirnov:2024onl, Belitsky:2023qho}, taking into account
that denominators have factorized dimension and kinematic dependence
in our reduction results \cite{Smirnov:2020quc,
  Usovitsch:2020jrk}. The \textsc{FIRE} program takes care both of the
reduction and the reconstruction, and can be used on HPC systems using
MPI for the task distribution.  While initially aimed for problems
with more variables, the balanced Zippel approach is also efficient
for problems with two variables, automatically exploiting sparsity of
the polynomials in kinematic variables.  The soft-expanded integrals
requiring IBP reduction has up to 13 powers of irreducible scalar
products and up to ten additional propagator powers.

\textbf{\textit{The radial action---}}The radial action for spinning amplitudes
can be obtained via the amplitude-action relation, schematically,
\begin{equation}
   i\mathcal M = |\boldsymbol p| \int d^{D-2} \boldsymbol b \ e^{i \boldsymbol b
     \cdot \boldsymbol q} (e^{i I_r} - 1) \;,
     \label{eq:aarelation}
\end{equation}
where $\boldsymbol b$ and $\boldsymbol q$ are the transverse spatial
parts of the impact parameter and transfer momentum respectively, and
$\boldsymbol{p}$ is the three-momentum at past infinity. Here $I_r$ is
the radial action, which is defined classically through the radial
momentum, $p_r$, along the scattering trajectory; see
Refs.~\cite{Bern:2021dqo,Bern:2021yeh,Bern:2023ccb,Barack:2023oqp,
  Bern:2024adl, Akpinar:2024meg} for more details. An equivalent
formulation, at least in the spinless case, is the exponential
representation of the S matrix proposed in
Refs.~\cite{Damgaard:2021ipf, Damgaard:2023ttc}. In the context of
perturbation theory, the amplitude and radial action are expanded in
powers of the coupling, $G$,
\begin{eqnarray}
  \label{eq:pexpansions}
  && \mathcal{M}=\sum_{k}G^k\mathcal{M}_{k}\;,\ \ {\tilde{I}}_r=\sum_{k}{G^k\tilde{I}}_{r,k}\;, \\ 
  && \quad \tilde{I}_r = |\boldsymbol p| \int d^{D-2} \boldsymbol b \ e^{i \boldsymbol b
             \cdot \boldsymbol q} I_r\;.
\end{eqnarray}
In this way, the first few orders of equation~(\ref{eq:aarelation}) read
\allowdisplaybreaks{
\begin{eqnarray}
  \mathcal{M}_1={}&&{\tilde{I}}_{r,1}\nonumber\;,\\
  \label{eq:aarelationPert2}
  \mathcal{M}_2={}&&{\tilde{I}}_{r,2}+\int_{\boldsymbol{\ell}}\frac{{\tilde{I}}_{r,1}{\tilde{I}}_{r,1}}{Z_1}\;,\\
  \label{eq:aarelationPert3}
  \mathcal{M}_3={}&&{\tilde{I}}_{r,3}+\int_{\boldsymbol{\ell}}\frac{{\tilde{I}}_{r,2}{\tilde{I}}_{r,1}}{Z_1}+\int_{\boldsymbol{\ell}}\frac{{\tilde{I}}_{r,1}{\tilde{I}}_{r,1}{\tilde{I}}_{r,1}}{Z_1Z_2}\nonumber\;,
\end{eqnarray}
}%
where $Z_i$ are three-dimensional propagators whose explicit forms are
not relevant here. The important point is that the lower-order
iterations precisely correspond to the superclassical divergences
coming from the amplitudes, such that the radial action is simply the
finite part of loop-level amplitudes. In practice, this is obtained
through two loops by simply dropping the double- and triple-iteration
ladderlike master integrals, as was done in the spinless case in
Ref.~\cite{Bern:2021xze} and the quadratic-in-spin case in
Ref.~\cite{Akpinar:2024meg}. We follow the proposal of
  Ref.~\cite{Gonzo:2024zxo} that a single radial action encodes
  complete information about scattering observables even beyond the
  aligned spin limit, and apply the covariant Dirac bracket
  formalism for spinning two-body dynamics developed in the same article
  to calculate such observables to $\mathcal O(G^3 S^4)$.

\begin{table*}
\centering
\begin{tabular}{|c|c|c|c|c|c|c|}
\hline
$\mathcal{O}_{(n,m)}$ \ \ & $m = 0$ & $m = 1$ & $m = 2$ & $m = 3$ & $m = 4$  & $m = 5$\\[0.15cm] 
\hline
$n = 0$ & $\mathds{1}$& -- & -- & -- & -- & --\\[0.15cm] 
$n = 1$ & $i (n \cdot S)$ & -- & -- & -- & -- & --\\[0.15cm] 
$n = 2$ & $\Omega_-$ & $\Omega_+$ & $q^2 (\bar{p}_2 \cdot S)^2$ & -- & -- & --\\[0.15cm] 
$n = 3$ & $i(n \cdot S)\Omega_-$ & $i(n \cdot S)\Omega_+$ & $iq^2(n \cdot S)(\bar{p}_2 \cdot S)^2$ & -- & -- & --\\[0.15cm] 
$n = 4$ & $\Omega_-^2$ & $\Omega_+^2$ & $\Omega_+\Omega_-$ & $q^2 (\bar{p}_2 \cdot S)^2\Omega_-$ & $q^2 (\bar{p}_2 \cdot S)^2\Omega_+$ & $q^4 (\bar{p}_2 \cdot S)^4$\\[0.15cm] 
\hline
\end{tabular}
\caption{Spin structures for the ansatz~\eqref{eq:spinningAnsatz}, with $n^\mu \equiv
\epsilon^{\mu\nu\alpha\beta}q_\nu\bar{p}_{1\alpha}\bar{p}_{2\beta}$ and the
$\Omega_\pm$ structures defined in equation~\eqref{eq:omega}.}
\label{tab:spinStructures} 
\end{table*}

\textbf{\textit{Resolving spin structures---}}The spin vector for a
massive bosonic spin-$s$ particle as an operator in spin space takes
the form
\begin{equation}
    \mathds{S}^\mu_{ij} = \frac{(-1)^s}{2m}\epsilon^{\mu \lambda \alpha \beta} p_\lambda \varepsilon^\star_{i I(s)}(p)\big(\Sigma_{\alpha \beta}\big)^{I(s)}_{~J(s)}\varepsilon^{J(s)}_j(p)\;,
\end{equation}
where $i,j$ are little group indices, 
$(-1)^s$ accounts for the normalization of the polarization tensors,
and where the Lorentz generators in the
spin-$s$ representation, with indices $I(s)$ and $J(s)$, are defined as
\begin{equation}
	\big(\Sigma_{\alpha \beta}\big)^{I(s)}_{~J(s)} = 2 i s \delta^{(I_1}_{~~[\alpha} \eta^{\vphantom{\rho}}_{\beta] (J_1}\delta^{I_2}_{~~J_2}\ldots \delta^{I_s)}_{~~J_s)}\;,
\end{equation}
where symmetrizations over $n$ indices includes a factor of $1/n!$.
To make a connection
with the scattering amplitudes, we define the spin vector in the frame
of an average momentum $\bar{p}^\prime_{1} = \bar{p}_1(m_1/\bar{m}_1)$
and relate $\varepsilon_4^{\star I}, \varepsilon_1^{J}$ to
polarizations $\bar{\varepsilon}^{\star I},\bar{\varepsilon}^{J}$ in
the aforementioned frame through Lorentz
transformations~\cite{Ross:2007zza,Vaidya:2014kza,Maybee:2019jus} (see Ref.~\cite{Akpinar:2024meg} for more details).
The classical spin vector then follows as an expectation value over all external spin states
\begin{equation}
  S^\mu = \sum_{i,j}\omega_i^\star \mathds{S}^\mu_{ij} \omega_j\;,
 \qquad \sum_i \omega_i^\star \omega_i = 1\;,
\end{equation}
where $\omega_i$ are arbitrary complex weights for polarizations of
different spin states.  Products of the classical spin vector are
mapped to symmetrized products of spin operators
\begin{equation}
	S^\mu S^\rho \ldots ~S^\sigma S^\nu   = \sum_{i,k,...}\omega_i^\star \mathds{S}^{(\mu}_{ik}\mathds{S}^{\rho}_{k\ell} \ldots \mathds{S}^{\sigma}_{rm} \mathds{S}^{\nu)}_{mj}\omega_j\;.
\end{equation}
To disentangle ambiguous spin-Casimir contributions to the scattering amplitudes
we employ the spin interpolation method introduced in
Ref.~\cite{Akpinar:2024meg} as follows.  Given that we know what spin structures
to expect in an amplitude involving a spin-$s$ particle, we can write an ansatz
valid for all representations that we consider, that is for $s=0, 1$ and $2$.
Then we interpret spin universality as the statement that the
coefficients appearing in the ansatz---classical or quantum---are independent
of the spin representation. The ansatz is naturally organized as
an expansion in $S$ and $q$. In this way, we can truncate the ansatz to a desired order in 
spin by retaining only terms up to the same order in $q$, as terms of higher orders in
spin are always accompanied by higher orders in $q$ due to the
classical power counting.

Therefore, we write an ansatz for the finite part of the amplitude
in terms of all possible spin structures that appear through
$\mathcal{O}(S^4)$
\begin{equation}
	\mathcal{M}_{\text{fin.}}^{(L)} = \frac{\kappa^{2+2L}}{(-q^2)^{1-L/2+L\epsilon}}\sum_{n = 0}^{4}\sum_{m } \alpha^{(L, n,m)} \mathcal{O}^{(n,m)}\;,
	\label{eq:spinningAnsatz}
\end{equation}
where the coefficients $\alpha^{(L,n,m)} $ contain both classical and quantum
contributions, $L$ denotes the number of loops, $n$ denotes the spin orders, and $m$
organizes each spin contribution within a given order. Furthermore, we include the
appropriate prefactor for an $L$-loop amplitude that is regularized using
dimensional regularization, $D = 4 -2 \epsilon$. 
The spin structures $\mathcal{O}^{(n,m)}$ are given in
Table~\ref{tab:spinStructures}.  To study the so-called spin-shift
symmetry~\cite{Bern:2022kto,Aoude:2022trd,Chen:2022clh,Akpinar:2024meg}
of the amplitude under the transformation $S^\mu\to S^\mu+\xi q^\mu$
for arbitrary $\xi$, the ansatz is written in a basis involving
\begin{equation}
 	\Omega_\pm = (q\cdot S)^2 \pm q^2 S^2\;,
 	\label{eq:omega}
\end{equation}
where $\Omega_-$ preserves the spin-shift symmetry, whereas $\Omega_+$ breaks it.

To fix the coefficients in the ansatz, we first evaluate
\eqref{eq:spinningAnsatz} in the $s = 0,1$ and $2$ representations and
expand to $\mathcal{O}(q^4)$. In the ansatz, this means writing the
$\alpha^{(L,n,m)}$ as an expansion in $q$. In this way, we find $14$ classical
coefficients---given by the leading $q$ contribution of each
structure in Table~\ref{tab:spinStructures}---as well as $6$
quantum-suppressed coefficients.  Then, we project onto the
appropriate form factor basis for each representation [see,
e.g.,~\eqref{eq:FFdec}], and solve a linear system of equations by
demanding equality with the corresponding spinning amplitude. This
matching requires that the amplitudes are also expanded to the same
order in $q$. In the end, we find a
system of $31$ linear relations, of which $11$ are linearly
dependent. This linear dependence is an explicit manifestation of spin
universality, since terms fixed by lower-spin amplitudes, such as
spinless terms and spin-orbit interaction terms, reappear in
higher-spin amplitudes. The $20$ linearly independent relations then
are the final product of our spin interpolation method, which we
use to compute the $20$ spin coefficients in terms of amplitude
coefficients. In the end we obtain a set of linear relations which
determine the ansatz coefficients to fourth-order in spin in terms of
the finite part of the $L$-loop amplitude coefficients in any theory.

With this we reproduce the known classical amplitudes in the
literature through $\mathcal{O}(G^2 S^4)$ and, for the first time,
obtain the classical amplitude to $\mathcal{O}(G^3 S^4)$. Our results
are explicitly given in the Supplemental Material~\cite{S4Results}. We
find the spin-shift symmetry to be present in the probe structures at
$\mathcal{O}(G^3)$ through fourth order in spin.  Since we have
extracted the classical amplitude, from now on we can ignore quantum
corrections, for example, in the difference between $\bar{m}_i$ and
$m_i$.

Then, using our classical spinning amplitude through two loops, the radial action follows as
\begin{equation}
 I_r = \int \frac{d^4 q}{(2\pi)^2}\, \delta(p_1 \cdot q) \delta(-p_2 \cdot q) 
 e^{i b \cdot q} \mathcal{M}_{\text{fin.}}\;.
 \label{eq:RadialActionFourierTransform}
\end{equation}
To validate our results for $I_r$, we find agreement with 
the direct classical calculation in Ref.~\cite{Gonzo:2024zxo} through
$\mathcal{O}(G^3 S^4)$ in the limit of a spinless probe in a Kerr
background.

\textbf{\textit{Observables---}}A new method to compute observables from the
radial action was recently proposed in Ref.~\cite{Gonzo:2024zxo}, which makes
use of covariant Dirac brackets. Here we show how these
brackets, together with the radial action obtained from the finite piece of the
amplitude, can be used to compute nonaligned spin observables. 

Let us first discuss the derivation of the brackets for the case of two
spinning massive particles following Ref.~\cite{Gonzo:2024zxo}. Although
our discussion so far has involved only one spinning black hole, we
will consider the more general case with two spinning black holes (at the
end, we can set one spin to zero).
Note that we work with normalized four-momenta, $u_i$, and spin
tensors $S^{\mu \nu}_i$.

The relevant dynamics for the radial action is in the asymptotic past, where we have 
straight line trajectories
\begin{equation}
    x_i^\mu(\tau_i) = u_i^\mu \tau_i + b_i^\mu\;, \quad i = 1,2\;,
\end{equation}
where $\tau_i$ is the proper time along the trajectory and $b_i$
is the initial trajectory. The nonvanishing Poisson brackets read
\begin{equation}
\begin{split}
	& \hspace{2cm} \{ b_i^\mu, u_i^\nu\}_{\text{P.B}} = -\eta^{\mu \nu}/m_i\;, \\
	& \{ S_i^{\mu \nu}, S_i^{\alpha \beta} \}_{\text{P.B}} = S_i^{\mu \alpha} \eta^{\nu \beta} - S_i^{\mu \beta} \eta^{\nu \alpha} - (\mu \leftrightarrow \nu) \;.
\end{split}
\end{equation}
Next we impose all second class constraints of the system, which
promotes Poisson brackets to Dirac brackets~\cite{Gonzo:2024zxo}. More
specifically, there are two spin-related
constraints known as the spin-supplementary and the conjugate gauge
conditions~
\begin{equation}
 V_i^\mu = u_{i \nu} S^{\mu \nu}_i = 0\;, 
 \quad \chi_i^\mu = \Lambda_{i,0}^{~\mu} - u_i^\mu = 0\;,
\end{equation}
where $\Lambda_0^\mu$ is the timelike component of the body-fixed frame (see Ref.~\cite{Hanson:1974qy} 
for more details). Moreover, there are four constraints following from transversality and on-shellness
\begin{equation}
    \begin{split}
        & \phi_1 = b \cdot u_1 = 0\;, \qquad \phi_2 = u_1^2 = 1\;, \\
        & \phi_3 = b \cdot u_2 = 0\;, \qquad \phi_4 = u_2^2 = 1\;,
    \end{split}
\end{equation}
where $b^\mu = b_2^\mu - b_1^\mu$ is the impact parameter. In this way, the
nonvanishing Dirac brackets read
\begin{eqnarray}
&&\hspace{-0.4cm}\{b^\mu, u_i^\nu\} = -\operatorname{sgn}_i \frac{(y^2-1)b^\mu b^\nu + l^\mu l^\nu}{m_i b^2 (y^2-1)}\;, \\
&&\hspace{-0.4cm}\{b^\mu, a_i^\nu\} = \operatorname{sgn}_i \frac{u_i^\nu ((y^2-1)b^\mu(b\cdot a_i) + l^\mu(l \cdot a_i))}{m_i b^2 (y^2-1)}\;, \\
&&\hspace{-0.4cm}\{b^\mu,b^\nu\} = \frac{(b^2(y u_2^\mu - u_1^\mu) - l^\mu(a_1 \cdot u_2))b^\nu - (\mu \leftrightarrow \nu)}{m_1b^2(y^2-1)} \nonumber \\
&&\hspace{-0.4cm}\hspace{2cm} + \ (1\leftrightarrow 2, b \rightarrow -b)\;, \\
&&\hspace{-0.4cm}\{a^\mu_i, a^\nu_i\} = \frac{\epsilon^{\mu \nu}_{~~\rho \sigma} u_i^\rho a_i^\sigma}{m_i}\;,
\end{eqnarray}
where $\operatorname{sgn}_1  = -1$ and $\operatorname{sgn}_2  = 1$. To simplify
the expressions we defined 
\begin{eqnarray}
	\quad l^\mu = \epsilon^{\mu}_{~ \nu \rho \sigma} b^\nu u_1^\rho u_2^\sigma\;, \quad a_i^\mu = \frac{1}{2m_i}\epsilon^{\mu}_{~\nu \rho \sigma}u_i^\nu S_i^{\rho \sigma}\;.
\end{eqnarray} 
Note that there are sign differences 
compared to Ref.~\cite{Gonzo:2024zxo}, which is a consequence of an opposite metric
signature alongside an additional sign in defining the spin vector. 

The power of these brackets becomes apparent when computing observables.  Given
a radial action, $I_r$, we may compute the change in an observable
via~\cite{Gonzo:2024zxo}
\begin{equation}
    \Delta \lambda^\mu = \sum_j \frac{1}{j!}\{\underbrace{I_r,\{I_r,\ldots,\{I_r}_{j \ \text{times}},\lambda^\mu\}\ldots\}\}\;,
    \label{eq:obsFormula}
\end{equation}
where $\lambda^\mu$ can be either the normalized four-momenta or spin
vectors.  Notice that this approach bypasses the need for explicit cut
diagrams, since these contributions appear in the iterative structures
of equation~\eqref{eq:obsFormula}. For our purposes, we simply set
$a_1 = a$ and $a_2 = 0$ when computing the impulse and spin kick
through $\mathcal{O}(G^3 S^4)$. In doing so, we find exact agreement
with known results at $\mathcal{O}(G^3 S^2)$ including both probe
structures and self-force
corrections~\cite{Jakobsen:2022fcj,FebresCordero:2022jts,
  Akpinar:2024meg}.  The results at $\mathcal{O}(G^3 S^3)$ and
$\mathcal{O}(G^3 S^4)$ are new and we provide them in the supplemental
material~\cite{S4Results}. We have also found agreement with the
post-Newtonian velocity expansion results for the scattering angle in
the aligned-spin limit through $O(G^3 S^4)$ \cite{Bautista:2024agp}.

We combine our result with those of Alessio and Di
Vecchia~\cite{Alessio:2022kwv} to find that the logarithmic divergence---appearing in the high-energy limit of the two-loop amplitude---cancels through fourth order in spin. The cancellation was previously
observed at $O(G^3)$ in the spinless case \cite{DiVecchia:2020ymx,
  Rothstein:2024nlq} and through the second order in spin
\cite{Jakobsen:2022zsx}. Moreover, treating their results as a
radiation-reaction contribution to the radial action---which is a
unique finite contribution at $O(G^3)$---we compute the impulse and
spin kick to all orders in spin, finding exact agreement with the
known results of Ref.~\cite{Jakobsen:2022zsx} to quadratic order in
spin. These results can be found in the Supplemental
Material~\cite{S4Results}.

\textbf{\textit{Conclusion---}}In this Letter we have computed for the
first time the classical two-loop amplitude for the conservative
scattering process between a spinning and a spinless black hole
through fourth order in spin. We have done this by computing
scattering amplitudes between a scalar massive particle and a spin-$s$
particle ($s=0,1$ and $2$) and using the recently introduced spin
interpolation method~\cite{Akpinar:2024meg} to extract all spin
structures in the classical amplitude. We find that the spin-shift
symmetry~\cite{Bern:2022kto,Chen:2022clh,Akpinar:2024meg} is present
in all probe structures. It is natural to conjecture that this is
related to a hidden symmetry, and hence integrability, of probe orbits
in a Kerr background through fourth order in spin of either the
background or probe object. This goes beyond the quadratic order in
spin where integrability of binary Kerr systems has already been
established~\cite{Carter:1968rr, rudiger1981conserved,
  Rudiger1983conserved, Page:2006ka, Compere:2023alp}.  To compute
observables, we use the amplitude-action relation, and we find
agreement with the radial action known in the literature for the case
of a probe scalar. We then compute non-aligned spin observables by
using the covariant Dirac brackets introduced in
Ref.~\cite{Gonzo:2024zxo}. With this we reproduce known observables to
$\mathcal{O}(G^3S^2)$, giving further evidence for
equation~\eqref{eq:obsFormula}.  To complete the nondissipative
picture, we have used the results of Ref.~\cite{Alessio:2022kwv} to
show the cancellation of the logarithmic divergence in the high-energy
limit, and we have further obtained observables for the
radiation-reaction dynamics to all orders in spin with the covariant
Dirac brackets. Though we have yet to rigorously establish the radial
action and Dirac bracket formalisms for nonaligned spin observables,
the various checks that we have performed provide strong evidence for
the correctness of the formalisms, likely to all orders, as the
two-loop case remains unchanged from the one-loop case. This provides
a simple and compelling answer to the quest of finding an eikonal-like
formula for spinning binary dynamics \cite{Bern:2020buy,
  Jakobsen:2021zvh, Cristofoli:2021jas, Luna:2023uwd, Gatica:2023iws}.

Our techniques can be further exploited in several directions.  We can
complete the two-loop picture for quartic-in-spin observables by
studying the scattering of two spinning black holes. We can also
extend the conservative picture by incorporating dissipative effects,
which requires including a small number of additional diagrams that
vanish in the potential region in our scattering amplitude calculation
as well as using the values of master integrals in the full soft
region. We would also like to go beyond the quartic order in spin by
matching with black hole perturbation theory~\cite{Bautista:2022wjf}
and study more general spinning objects such as neutron stars. Another
important direction would be to explore higher orders in $G$.
Finally, it would be interesting to explicitly connect the spin-shift
symmetry observed here with the integrability of probe orbits in Kerr
backgrounds, e.g.,~by recasting the symmetry of the amplitude as a
certain classical Ward identity, and this could potentially offer a
novel on-shell amplitude perspective on hidden symmetries of classical
systems.

\textbf{\textit{Acknowledgements---}}We thank Rafael Aoude, Fabian
Bautista, Lucile Cangemi, and Riccardo Gonzo for many insightful
discussions. We are especially thankful to Graham R. Brown for help
with confirming the agreement of some of our results with the
literature.
The Mathematica package \texttt{FeynCalc}
\cite{Shtabovenko:2023idz} was used extensively in some of the calculations.
D.A.\ is supported by a STFC studentship.
The work of F.F.C. is supported in part by the U.S. Department of Energy
under grant DE-SC0010102.
M.K.\ is supported by the DGAPA-PAPIIT grant IA102224 (``Iluminando agujeros
negros") and the PIIF at UNAM.
This work has been made possible in part through the support of the FSU
Council on Research and Creativity (``Black Holes Under the Microscope";
SEED Grant, 2023).
A.S. is supported in part by the Ministry of Education and Science of the Russian Federation
as part of the program of the Moscow Center for Fundamental and Applied Mathematics
under Agreement No. 075-15-2025-345.
M.Z.’s work is supported in part by the U.K.\ Royal Society through
Grant URF\textbackslash R1\textbackslash 20109.
The study of A.S. was conducted under the state assignment of Lomonosov Moscow State University.
For the purpose of open access, the authors
have applied a Creative Commons Attribution (CC BY) license to any
Author Accepted Manuscript version arising from this submission.
This research used resources of the National Energy Research Scientific
Computing Center (NERSC), a Department of Energy Office of Science User Facility
using NERSC award HEP-ERCAP m4394 for 2024 and 2025.

\bibliography{spin2.bib}

\newpage
\appendix
\onecolumngrid
\vspace{0.8cm}
\begin{center}
\textbf{Supplemental Material}
\end{center}
Here we present full results for the classical spinning amplitudes that we have
computed by providing the corresponding coefficients of the ansatz of
equation~\eqref{eq:spinningAnsatz}. In the following results, we homogenize the
mass dependence in all spin structures by using $u_i = \bar{p}_i/\bar{m}_i$ and
$a = S/\bar{m}_1$, and we drop the barred notation since $\bar{m}_i = m_i +
\mathcal{O}(q)$.

\vspace{4mm}
\begin{center}
\text{\textit{Tree-level Amplitude}}
\end{center}
The tree-level amplitude is well known and takes a very simple form
\begin{equation}
	\begin{split}
		\mathcal{M}^{(0)} = \frac{\kappa^2}{(-q^2)}\Big\{& (2y^2-1) m_1 m_2 + i y m^2_1 m^2_2 (n \cdot a) + \frac{(2y^2-1) m^2_1 m^2_2}{4}(q \cdot a)^2 \\
		& + \frac{i y m^2_1 m^2_2}{6}(n \cdot a)(q \cdot a)^2 + \frac{(2y^2-1) m^2_1 m^2_2}{48}(q \cdot a)^4\Big\},
	\end{split}
\end{equation}
where we have dropped terms that are proportional to $q^2$ as they cancel the
$t$-channel pole.

\vspace{4mm}
\begin{center}
\text{\textit{One-loop Amplitude}}
\end{center}
The one-loop amplitude takes the form of equation~\eqref{eq:spinningAnsatz} with
coefficients
\allowdisplaybreaks{
\begin{eqnarray}
	&& \alpha^{(1,0,0)} = \frac{3m_1^2 m_2^2(m_1 + m_2)}{512}(5y^2-1)\;, \quad \alpha^{(1,1,0)} = \frac{m_1^2 m_2^2(4 m_1 + 3 m_2)}{512(y^2-1)}y(5y^2-3)\;, \nonumber\\
	&& \alpha^{(1,2,0)} = \frac{m_1^2 m_2^2}{4096 (y^2-1)}((95 y^4 - 102 y^2 + 15) m_1 +  4 (15 y^4 - 15 y^2 + 2) m_2)\;, \quad \alpha^{(1,2,1)} = 0\;, \nonumber\\ 
	&& \alpha^{(1,2,2)} = -\frac{m_1^2 m_2^2}{2048 (y^2-1)^2}((65 y^4 - 66 y^2 + 9) m_1 +  2 (15 y^4 - 12 y^2 + 1) m_2)\;, \nonumber\\
	&& \alpha^{(1,3,0)} = \frac{m_1^2 m_2^2}{1024 (y^2-1)}y((9 y^2 - 5) m_1 + (5 y^2 -2) m_2)\;, \quad \alpha^{(1,3,1)} = 0\;, \nonumber\\
	&& \alpha^{(1,3,2)} = -\frac{m_1^2 m_2^2}{1024 (y^2-1)^2}y(8(2y^2 - 1) m_1 + (5 y^2 + 1) m_2)\;, \\
	&& \alpha^{(1,4,0)} = \frac{m_1^2 m_2^2}{98304 (y^2-1)}((239 y^4- 250 y^2+35 ) m_1 + 24 y^2 (5 y^2 - 4) m_2)\;, \quad \alpha^{(1,4,1)} = \alpha^{(1,4,2)} = 0\;, \nonumber\\
	&& \alpha^{(1,4,3)} = -\frac{m_1^2 m_2^2}{24576 (y^2-1)^2}((193 y^4 - 194 y^2 + 25) m_1 + 12 (5 y^4 - y^2 - 2)m_2)\;, \quad \alpha^{(1,4,4)} = 0\;, \nonumber\\
	&& \alpha^{(1,4,5)} = \frac{m_1^2 m_2^2}{12288 (y^2-1)^3}(8 (8 y^4 - 8 y^2 + 1) m_1 + (15 y^4 + 6 y^2 -13) m_2)\nonumber\;,
\end{eqnarray}
}%
which is in exact agreement with known results, see for example
Refs.~\cite{Aoude:2022trd,Chen:2021qkk,Bern:2022kto}.

\vspace{4mm}
\begin{center}
\text{\textit{Two-loop Amplitude}}
\end{center}
Finally, the two-loop amplitude takes the form of 
equation~\eqref{eq:spinningAnsatz} where we further organize the coefficients as 
\begin{equation}
	\alpha^{(2,n,m)} = \alpha^{(2,n,m)}_{\text{probe}} + m_1^3 m_2^3 \, \alpha^{(2,n,m)}_{\text{1SF}}.
\end{equation}
In such way, the probe contributions read
\allowdisplaybreaks{
\begin{eqnarray}
	&& \alpha_{\text{probe}}^{(2,0,0)} = \frac{
	m_1^2 m_2^2 \left(m_1^2+m_2^2\right) \left(64 y^6-120 y^4+60 y^2-5\right)}{49152 \pi ^2 \epsilon \left(y^2-1\right)^2} \;,\nonumber \\
	&& \alpha_{\text{probe}}^{(2,1,0)} = \frac{m_1^2m_2^2(3m_1^2+2m_2^2)y \left(16 y^4-20 y^2+5\right)}{24576 \pi ^2 \epsilon \left(y^2-1\right)^2 }\;, \nonumber\\
	&& \alpha_{\text{probe}}^{(2,2,0)} = \frac{m_1^4 m_2^2 \left(128 y^6-216 y^4+96 y^2-7\right)+ m_1^2 m_2^4 \left(64 y^6-104 y^4+44 y^2-3\right)}{98304 \pi ^2\epsilon  \left(y^2-1\right)^2 }\;, \nonumber\\
	&& \alpha_{\text{probe}}^{(2,2,1)} = 0 \;,\nonumber \\ 
	&& \alpha_{\text{probe}}^{(2,2,2)} = -\frac{m_1^4 m_2^2 \left(80 y^6-132 y^4+57 y^2-4\right)+ m_1^2 m_2^4 \left(32 y^6-48 y^4+18 y^2-1\right)}{49152 \pi ^2\epsilon \left(y^2-1\right)^3 }\;, \nonumber\\
	&& \alpha_{\text{probe}}^{(2,3,0)} = y \frac{ m_1^4 m_2^2\left(80 y^4-92 y^2+21\right)+2  m_1^2 m_2^4\left(16 y^4-16 y^2+3\right)}{147456 \pi ^2 \epsilon \left(y^2-1\right)^2 }\;, \\
	&& \alpha_{\text{probe}}^{(2,3,1)} = 0\;, \nonumber\\
	&& \alpha_{\text{probe}}^{(2,3,2)} = -\frac{ m_1^4 m_2^2y \left(64 y^4-70 y^2+15\right)+2 m_1^2 m_2^4 y^3 \left(8 y^2-5\right)}{73728 \pi ^2 \epsilon \left(y^2-1\right)^3 }\;, \nonumber\\
	&& \alpha_{\text{probe}}^{(2,4,0)} = \frac{3  m_1^4 m_2^2\left(64 y^6-104 y^4+44 y^2-3\right)+m_1^2 m_2^4 \left(64 y^6-88 y^4+28 y^2-1\right)}{1179648 \pi ^2 \epsilon \left(y^2-1\right)^2 }\;, \nonumber\\
	&& \alpha_{\text{probe}}^{(2,4,1)} = 0\;, \nonumber\\
	&& \alpha_{\text{probe}}^{(2,4,2)} = 0\;, \nonumber\\
	&& \alpha_{\text{probe}}^{(2,4,3)} = -\frac{3 m_1^4 m_2^2\left(48 y^6-76 y^4+31 y^2-2\right)+ m_1^2m_2^4 \left(32 y^6-32 y^4+2 y^2+1\right)}{294912 \pi ^2 \epsilon \left(y^2-1\right)^3 }\;, \nonumber\\
	&& \alpha_{\text{probe}}^{(2,4,4)} = 0\;, \nonumber\\
	&& \alpha_{\text{probe}}^{(2,4,5)} = \frac{3m_1^4  m_2^2 \left(128 y^6-200 y^4+80 y^2-5\right)+m_1^2 m_2^4 \left(64 y^6-40 y^4-20 y^2+5\right)}{1179648 \pi ^2 \epsilon \left(y^2-1\right)^4 }\;,\nonumber
\end{eqnarray}
}%
where the vanishing coefficients directly imply that the amplitude in both probe
limits satisfies spin-shift symmetry. On the other hand, the non-probe, 1SF
contributions read
\allowdisplaybreaks{
\begin{eqnarray}
	&& \alpha_{\text{1SF}}^{(2,0,0)} = \frac{y \left(36 y^6-114 y^4+132 y^2-55\right)}{24576 \pi ^2  \epsilon \left(y^2-1\right)^2} + \frac{\left(4 y^4-12 y^2-3\right) \operatorname{arcosh}(y)}{4096 \pi ^2  \epsilon \sqrt{y^2-1}}\;, \nonumber\\
	&& \alpha_{\text{1SF}}^{(2,1,0)} = \frac{\left(36 y^6-156 y^4+84 y^2+41\right)}{24576 \pi ^2 \epsilon \left(y^2-1\right)^2 } + \frac{y \left(2 y^4-11 y^2-6\right) \operatorname{arcosh}(y)}{2048 \pi ^2  \epsilon \left(y^2-1\right)^{3/2}}\;, \nonumber\\
	&& \alpha_{\text{1SF}}^{(2,2,0)} = \frac{y \left(366 y^6-2499 y^4+550 y^2+1218\right)}{491520 \pi ^2 \epsilon \left(y^2-1\right)^2 } + \frac{\left(16 y^8-144 y^6+13 y^4+78 y^2+12\right) \operatorname{arcosh}(y)}{32768 \pi ^2 \epsilon \left(y^2-1\right)^{5/2}}\;, \nonumber\\ 
	&& \alpha_{\text{1SF}}^{(2,2,1)} = -\frac{y \left(2 y^6-5 y^4+114 y^2+264\right)}{163840 \pi ^2 \epsilon  \left(y^2-1\right)^2} -\frac{3\left(3 y^4+18 y^2+4\right) \operatorname{arcosh}(y)}{32768 \pi ^2  \epsilon \left(y^2-1\right)^{5/2}}\;, \\
	&& \alpha_{\text{1SF}}^{(2,2,2)} = -\frac{y \left(138 y^6-1769 y^4+1394 y^2+2122\right)}{245760 \pi ^2 \epsilon \left(y^2-1\right)^3 } -\frac{\left(8 y^8-104 y^6+35 y^4+162 y^2+24\right) \operatorname{arcosh}(y)}{16384 \pi ^2  \epsilon \left(y^2-1\right)^{7/2}}\;, \nonumber\\
	&& \alpha_{\text{1SF}}^{(2,3,0)} = \frac{\left(372 y^8-3524 y^6-653 y^4+1945 y^2-30\right)}{1474560 \pi ^2 \epsilon \left(y^2-1\right)^3}+\frac{y \left(16 y^8-184 y^6-50 y^4+57 y^2+35\right) \operatorname{arcosh}(y)}{98304 \pi ^2 \epsilon \left(y^2-1\right)^{7/2}}\;, \nonumber\\
	&& \alpha_{\text{1SF}}^{(2,3,1)} = -\frac{\left(4 y^8-12 y^6+385 y^4+1305 y^2+208\right)}{491520 \pi ^2 \epsilon \left(y^2-1\right)^3}-\frac{y \left(10 y^4+75 y^2+41\right)\operatorname{arcosh}(y)}{32768 \pi ^2 \epsilon \left(y^2-1\right)^{7/2}}\;, \nonumber\\
	&& \alpha_{\text{1SF}}^{(2,3,2)} = -\frac{\left(96 y^8-3724 y^6+2687 y^4+12809 y^2+1362\right)}{737280 \pi ^2 \epsilon\left(y^2-1\right)^4}-\frac{y \left(8 y^8-188 y^6-10 y^4+741 y^2+331\right) \operatorname{arcosh}(y)}{49152 \pi ^2 \epsilon \left(y^2-1\right)^{9/2}}\;, \nonumber\\
	&& \alpha_{\text{1SF}}^{(2,4,0)} = \frac{y \left(10596 y^{10}-139906 y^8+50884 y^6+120462 y^4-38972 y^2+25601\right)}{165150720 \pi ^2 \epsilon \left(y^2-1\right)^4} \nonumber\\ 
	&& \hspace{2cm}+ \frac{\left(64 y^{12}-960 y^{10}+314 y^8+280 y^6+534 y^4-14 y^2+55\right) \operatorname{arcosh}(y)}{1572864 \pi ^2 \epsilon \left(y^2-1\right)^{9/2}}\;, \nonumber\\
	&& \alpha_{\text{1SF}}^{(2,4,1)} = -\frac{y \left(4 y^{10}-18 y^8-108 y^6+1022 y^4+13732 y^2+11093\right)}{55050240 \pi ^2 \epsilon \left(y^2-1\right)^4} + \frac{\left(2 y^8-8 y^6-234 y^4-446 y^2-49\right) \operatorname{arcosh}(y)}{1572864 \pi ^2 \epsilon \left(y^2-1\right)^{9/2}}\;, \nonumber\\
	&& \alpha_{\text{1SF}}^{(2,4,2)} = -\frac{y \left(12 y^{10}-54 y^8+1636 y^6+6126 y^4-4604 y^2-2591\right)}{3932160 \pi ^2 \epsilon \left(y^2-1\right)^4} -\frac{\left(118 y^8+1064 y^6-90 y^4-934 y^2-53\right) \operatorname{arcosh}(y)}{786432 \pi ^2 \epsilon \left(y^2-1\right)^{9/2}}\;, \nonumber\\
	&& \alpha_{\text{1SF}}^{(2,4,3)} = -\frac{y \left(3312 y^{10}-129464 y^8+146636 y^6+657312 y^4-341941 y^2-157250\right)}{41287680 \pi ^2 \epsilon \left(y^2-1\right)^5} \nonumber\\ 
	&& \hspace{2cm} -\frac{\left(32 y^{12}-864 y^{10}+312 y^8+4944 y^6+450 y^4-3039 y^2-134\right) \operatorname{arcosh}(y)}{393216 \pi ^2 \epsilon \left(y^2-1\right)^{11/2}}\;, \nonumber\\
	&& \alpha_{\text{1SF}}^{(2,4,4)} = \frac{y \left(8 y^{10}-36 y^8+1156 y^6+1624 y^4-15747 y^2-6850\right)}{2752512 \pi ^2 \epsilon \left(y^2-1\right)^5} + \frac{\left(20 y^8+128 y^6-474 y^4-581 y^2-38\right) \operatorname{arcosh}(y)}{131072 \pi ^2 \epsilon \left(y^2-1\right)^{11/2}} \;, \nonumber\\
	&& \alpha_{\text{1SF}}^{(2,4,5)} = \frac{y \left(980 y^{10}-98474 y^8+122716 y^6+626512 y^4-602806 y^2-267223\right)}{41287680 \pi ^2 \epsilon \left(y^2-1\right)^6} \nonumber \\
	&& \hspace{2cm} +\frac{\left(16 y^{12}-624 y^{10}+74 y^8+5048 y^6-1474 y^4-4860 y^2-259\right) \operatorname{arcosh}(y)}{393216 \pi ^2 \epsilon \left(y^2-1\right)^{13/2}} \;. \nonumber
\end{eqnarray}
}

\end{document}